\documentclass{myclass}
\usepackage{multirow}
\usepackage[table]{xcolor}

\interspeechcameraready

\title{A Silent Speech Decoding System from EEG and EMG with Heterogenous Electrode Configurations}
\subtitle{Accepted for presentation at Interspeech 2025}
\author{Masakazu}{Inoue}
\author{Motoshige}{Sato}
\author{Kenichi}{Tomeoka}
\author{Nathania}{Nah}
\author{Eri}{Hatakeyama}
\author{Kai}{Arulkumaran}
\author{Ilya}{Horiguchi}
\author{Shuntaro}{Sasai}

\affiliation[nocounter]{}{Araya Inc.}{Japan}
\email{\{inoue,sato\_motoshige,tomeoka\_kenichi,nathania\_nah,hatakeyama\_eri,\\kai\_arulkumaran,horiguchi\_ilya,sasai\_shuntaro\}@araya.org}

\usepackage{comment}
\usepackage{subcaption}

\begin{document}

\maketitle

\begin{abstract}    
     Silent speech decoding, which performs unvocalized human speech recognition from electroencephalography/electromyography (EEG/EMG), increases accessibility for speech-impaired humans. However, data collection is difficult and performed using varying experimental setups, making it nontrivial to collect a large, homogeneous dataset. In this study we introduce neural networks that can handle EEG/EMG with heterogeneous electrode placements and show strong performance in silent speech decoding via multi-task training on large-scale EEG/EMG datasets. We achieve improved word classification accuracy in both healthy participants (95.3\%), and a speech-impaired patient (54.5\%), substantially outperforming models trained on single-subject data (70.1\% and 13.2\%). Moreover, our models also show gains in cross-language calibration performance. This increase in accuracy suggests the feasibility of developing practical silent speech decoding systems, particularly for speech-impaired patients.
\end{abstract}

\section{Introduction}
    
Silent speech decoding from electroencephalography (EEG) and electromyography (EMG) signals offers a promising communication solution for individuals with conditions like amyotrophic lateral sclerosis (ALS) or post-laryngectomy status.
While invasive brain activity measurements have achieved remarkable decoding accuracy \cite{willett2023high, anumanchipalli2019speech, metzger2023high, card2024accurate,moses2021neuroprosthesis}, their widespread adoption is hindered by the requirement for cranial surgery.

Conversely, non-invasive methods such as EEG and EMG are more accessible alternatives. Whilst there is considerable research on EEG-based speech decoding, these approaches have not yet achieved practical accuracy levels due to lower signal-to-noise ratios and spatial resolution, as compared to invasive methods \cite{defossez2023decoding,lee2020eeg,lee2023towards,liu2024eeg2text}.
For individuals with sufficient muscle control, EMG-based silent speech decoding has demonstrated adequate accuracy for silent speech command input and conversation, using subject-specific trained models \cite{meltzner2017silent,schultz2010modeling,wang2021all,benster2024}.
Considering that patients with neurodegenerative diseases may retain some degree of muscle control depending on their disease progression, it is practical to explore models that utilize EEG and EMG signals to achieve optimal accuracy.

However, there exists a lack of standardized equipment, as there is a variety of EEG headsets available with differing numbers of electrodes and variable placements, and EMG electrodes must be placed manually.
While achieving high accuracy in deep learning is known to require substantial amounts of data \cite{kaplan2020scaling, ye2024neural, sato2024scaling}, building large EEG datasets specifically designed for a particular task is challenging due to these variations in electrode placement protocols and measurement devices.
Therefore, developing architectures capable of handling diverse electrode arrangements is crucial for building generalizable EEG/EMG encoders.

Recent studies have made significant progress in developing generalizable decoders by using EEG data from diverse electrode configurations to train deep neural networks (DNNs) through self-supervised learning approaches, such as signal reconstruction \cite{chen2024eegformer,wang2023brainbert,jiang2024large}. However, the variability in shapes of these latents due to different electrode configurations remains, making it challenging to train a single model on diverse electrode configuration data in an end-to-end manner for downstream tasks. This has been addressed by incorporating average pooling \cite{peterson2021generalized,Han_2023} or implementing subject/electrode-specific layers \cite{spatial_attention,mentzelopoulos2024neural} to convert the latents into predefined sizes. 

While these studies demonstrated the utility of handling heterogeneous electrode configurations within a single model to increase available data for tasks such as motion/imagined motion detection, the effectiveness for relatively complex tasks such as silent speech decoding remains unclear, as does the efficacy of knowledge transfer between healthy individuals and patients. Therefore, in this work we investigate four types of EEG/EMG tokenizers, including a novel on-the-fly kernel, that enable our DNN models to handle data from heterogeneous electrode configurations. We evaluate how training models with these tokenizers on EEG/EMG data collected from eight healthy participants and a patient with a neurodegenerative disease during silent speech, along with additional pretraining on large data with heterogeneous electrode configurations and tasks, would improve silent speech decoding performance. Our results demonstrate higher word classification accuracy compared to models trained exclusively on individual data. Furthermore, we validate calibration performance on silent speech data for novel subjects, sessions from different days, and on a separate EMG dataset in a different language \cite{gaddy-klein-2020-digital}, demonstrating the value of pretraining on large datasets with heterogeneous tasks and electrode placements.

\begin{figure*}[!t]
  \centering
  \includegraphics[width=0.95\textwidth]{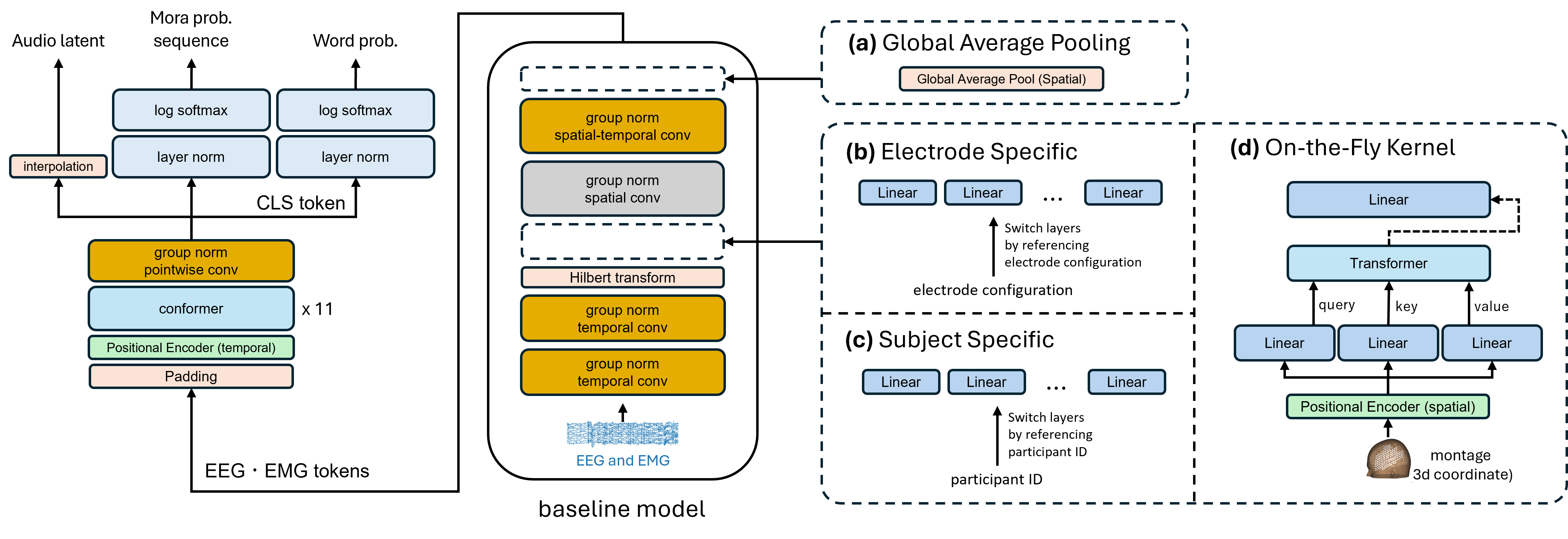}
  \caption{Our model architecture. To handle heterogeneous electrode configurations, the tokenizer can be selected by inserting with one of the following modules: (a) global average pooling, (b) electrode-specific, (c) subject-specific, or (d) on-the-fly kernel.}
  \label{fig:architecture}
\end{figure*}
\vspace{-5pt}

\section{Materials and methods}

\subsection{Data acquisition}\label{subsec:data_aquisition}

In previous work, \cite{sato2024scaling} collected a large EEG and EMG dataset recorded while participants spoke, using three types of EEG recording devices: eego sports\footnote{https://www.ant-neuro.com/products/eego-sports}, g.Pangolin\footnote{https://www.gtec.at/product/g-pangolin-electrodes/
} and g.SCARABEO\footnote{https://www.gtec.at/product/g-scarabeo-eeg-electrodes/}.

Using a similar setup, we collected EEG and EMG data with the eego sports device from eight healthy participants during silent speech and vocalized speech, and one patient during silent speech. Due to a progressive neuromuscular disease, the patient is unable to perform sufficient muscular movements for speech and cannot produce any vocalization due to dependence on a ventilator. 
For healthy participants, based on electrode positioning in \cite{gaddy-klein-2020-digital}, we recorded the EMG from the cheek, upper mouth, lower mouth, chin, and throat. For the patient, the latter EMG electrode pair was located below and above the eye instead of the throat because the electrode could not be attached to his throat due to the presence of medical equipment.
Participants were instructed to read text presented on a screen according to timing cues. Each healthy participant performed 20 sessions of word utterances and 2 sessions of sentence utterances in both vocalized and silent conditions. The patient also performed 20 sessions of silent speech word utterances, though some sessions were concluded prematurely due to fatigue.

The vocabulary size for the word reading task was 64 words, selected by the patient based on daily communication needs, and sentence reading sessions consisted of 153 sentences.
All participants are native Japanese speakers and spoke Japanese in our experiments. As such, to measure phonetic transcription accuracy, we use moras, which are Japanese phonetic units of speech \cite{kubozono2017mora}.

\begin{table*}[t]
  \caption{Pretraining datasets. For tasks, V+SS indicates paired vocalized/silent speech data, and VS indicates solely vocalized speech.
  The dataset ``word'' refers to subject-specific splits, and ``word all'' refers to the aggregation across all subjects.
  }
  \label{tab:datasets}
  \begin{minipage}{\textwidth}
  \centering
  \scalebox{0.77}{
  \begin{tabular}{ccccccccccc}
    \toprule
    \textbf{Task} &\multicolumn{2}{c}{\textbf{EEG}}&\textbf{EMG}&\textbf{Sampling Rate} & \textbf{Hours} & \textbf{Vocab Size}&\textbf{Confounds}& \multicolumn{3}{c}{\textbf{Dataset}}\\
    \arrayrulecolor{black!50}\midrule
    & \textbf{Device}  & \textbf{Channels}   &\textbf{Channels} & & & & & \textbf{word/word all} & \textbf{eego all} & \textbf{all}\\
    \arrayrulecolor{black}\midrule
    VS \cite{sato2024scaling} & g.Pangolin  & 128 ch & 3 ch &\SI{1024}{\hertz} & 49.1 & open &  audio  &  &  & \checkmark \\
    VS \cite{sato2024scaling} & g.Pangolin  & 128 ch & 3 ch &\SI{1024}{\hertz} & 64.2 & open &  audio, text  &  &  & \checkmark \\
    VS \cite{sato2024scaling} & g.SCARABEO  & 62 ch & 3 ch &\SI{1024}{\hertz} & 27.7 & open &  audio  &  &  & \checkmark  \\
    V+SS (word) & eego sports & 63 ch & 5 ch$^{*}$ &\SI{2048}{\hertz} & 16.1 & 64 & audio, text, timing, word & \checkmark & \checkmark & \checkmark\\
    V+SS (sentence) & eego sports & 63 ch & 5 ch$^{*}$ &\SI{2048}{\hertz} & 13.6 & 267 &  audio, text, timing  &  & \checkmark & \checkmark \\
    VS & eego sports & 63 ch & 3 ch &\SI{2048}{\hertz} & 49.2 & open & audio  &  & \checkmark & \checkmark \\
    \bottomrule
  \end{tabular}
  }
  \small 
  \end{minipage}
  \hspace*{\fill} \footnotesize{$^{*}$ EMG electrode placement differs between healthy participants and the patient (see Section~\ref{subsec:data_aquisition}). }
\end{table*}

\subsection{Model architecture}
We construct a baseline model combining HTNet \cite{lawhern2018eegnet,peterson2021generalized} and Conformer \cite{gulati2020conformer} as shown in Figure~\ref{fig:architecture} (left). The input signal undergoes initial processing through a tokenizer that operates on temporal and spatial dimensions, reducing the spatial dimension to one while converting the signal into EEG and EMG tokens on a sample-by-sample basis. These tokens are then batched and passed through a temporal positional encoder and are prepended with a CLS token before entering an 11-layer Conformer, which transforms them into EEG and EMG latents. These latents are then utilized by task-specific head modules for inference.

To accommodate multiple electrode configurations within a single model, we implement four alternative tokenization approaches to replace the base tokenizer, as shown in Figure \ref{fig:architecture} (right): (a) global average pooling tokenizer (GAP) that averages across electrode coordinates and compresses the spatial dimensions \cite{peterson2021generalized,Han_2023}; (b) electrode-specific tokenizer (ES) utilizing pre-configured electrode-specific linear layers \cite{spatial_attention}; (c) subject-specific tokenizer (SS) utilizing pre-configured subject-specific linear layers \cite{mentzelopoulos2024neural}; and (d) on-the-fly kernel tokenizer (OTFK) that transforms electrode dimensions to predefined dimensions adaptively.

\subsection{Electrode/subject-specific linear layers}
Models with the ES/SS tokenizers maintain predetermined sets of linear layers corresponding to the number of electrode/subject configurations, switching between appropriate linear layers based on electrode configuration or subject ID to transform electrode dimensions into predefined dimensions.

\subsection{On-the-fly kernel}
To handle datasets with heterogeneous electrode configurations within a single model, we propose an architecture that internally infers the weight of a linear layer (on-the-fly kernel) that transforms the spatial dimensions of electrode-based inputs to a predetermined number of spatial dimensions, based on electrodes' coordinates. Given $C$, the total number of EEG and EMG electrodes, and $K$, the predetermined size of the spatial dimension, the on-the-fly kernel inference module comprises of a positional encoder (PE) that embeds the 3D electrode coordinates into $K$ dimensions, linear layers for generating query, key, and value representations, and a single Transformer layer. 

For the PE, we use radial basis functions (with 9 kernels with variances $\sigma = \{\frac{2^{k-1}}{170}\ |\ k \in \{1, \ldots, 9\}\}$) to flexibly capture the spatial information of each electrode location, followed by a linear projection \cite{mentzelopoulos2024neural}. Given the three-dimensional coordinates $E\in\mathcal{R}^{C\times 3}$ for $C$ electrodes, the on-the-fly kernel $W$ is computed as follows:
\begin{align}
P&\in\mathcal{R}^{C\times K}  = \rm{PE}(E),\\
W'&= \mathrm{softmax}\left(\frac{\mathrm{FC_{query}}(P)\mathrm{FC_{key}}(P)^T}{\sqrt{K}}\right)\mathrm{FC_{value}}(P),\\
W&\in\mathcal{R}^{C\times K} = \mathrm{FC_{proj}}\left(W'\right),
\end{align}
where $\mathrm{FC_{query}}, \mathrm{FC_{key}},\mathrm{FC_{value}}$ and $\mathrm{FC_{proj}}$ are linear layers.

\subsection{Objective functions}

We employ multiple objective functions depending on the available confounds. For samples with audio latents, we calculate the mean squared error (MSE) loss between the EEG/EMG latents and the audio latents. For samples with word labels, we compute the cross-entropy loss between the predicted logits from the CLS token and the true word labels. For samples with transcribed speech text ground truth, we apply the connectionist temporal classification (CTC) loss \cite{graves2006connectionist} to the predicted mora sequence logits. For samples with ground truth moras for each time frame, we calculated the cross-entropy loss between the predicted mora logits and the true mora labels at each frame.

\section{Experiments}
\subsection{Preprocessing}

EEG data was preprocessed using a \SI{50}{\hertz} notch filter, common average reference, and 2-\SI{120}{\hertz} bandpass filter, followed by resampling to \SI{240}{\hertz}. EMG data recorded by bipolar electrodes underwent similar preprocessing, except for the common average reference. For patient data, an additional \SI{27}{\hertz} notch filter was applied to eliminate medical device interference. Audio captured from the microphone was downsampled to \SI{16}{\kilo\hertz} and converted to audio latents using wav2vec2.0 \cite{baevski2020wav2vec}.

\subsection{Training on heterogeneous EEG and EMG datasets}
We investigate how increasing the available training data size would enhance the robustness of word decoding performance for both healthy subjects and patients. Four datasets were employed in this study (Table~\ref{tab:datasets}):
\begin{itemize}
    \item ``word'': Single-subject word utterance task data used for training the baseline model (1 hour for patient, 1.9 hours for each healthy participant)
    \item ``word all'': all data of word utterance tasks from all subjects with two types of electrode configurations (16.1 hours) 
    \item ``eego all'': all data recorded using the eego sports device with three types of electrode configurations (76.9 hours)
    \item ``all'': all data listed in Table \ref{tab:datasets} with five types of electrode configurations (220 hours)
\end{itemize}
Each participant's data from the silent speech (word) task was split such that sessions 1-16 were used for training, sessions 17-18 for validation, and sessions 19-20 for test, ensuring equal word occurrence frequencies across splits.

Model training was performed using the AdamW optimizer \cite{loshchilov2017decoupled} for 300 epochs, with a learning rate starting at 3e-4 after an 8-epoch warmup period, gradually decreasing using a cosine scheduler. Model evaluation was based on word classification accuracy (WCA) of the silent speech (word) decoding task.

All models have $\sim$18.5 million parameters and were trained on an NVIDIA H100. Training took 2 hours per split for ``word'', 12 hours for ``word all'', 2 days for ``eego all'' and 4 days for the ``all'' dataset. For samples with an average of 3.3 \unit{\s}, the inference speed averaged 57 \unit{\ms}.
 
\subsection{Calibration for a novel healthy participant and new-day patient data}\label{subsec:calibration_jp}
We investigate the relationship between calibration performance and pretraining dataset size by calibrating models to a novel participant and new-day patient data. The models were initialized with weights pretrained on the `all', `eego-all', and `word-all' datasets.

For the novel healthy participant, we collected 15 sessions of EEG, EMG, audio, text, timing, and word data during silent speech (word) tasks. We trained 12 models by incrementally increasing calibration data from sessions 1 to 12. Session 13 was used for validation, and sessions 14-15 for test.

We collected additional patient data over three separate days with intervals of at least one week, recording 6 to 7 sessions per day. We used the first 3 and last 2 sessions for training and testing, with remaining session(s) for validation. During calibration, we trained our model by incrementally adding new-day sessions to the 16 sessions used for pretraining.

Models were trained using the AdamW optimizer over 100 epochs, with learning rate starting at 3e-4 after an 8-epoch warmup period and decreasing via cosine scheduler. Performance was evaluated using WCA on test data for each day.

\subsection{Calibration on a single-subject EMG English dataset}
We evaluate how our models, pretrained on our ``all'' dataset, could provide benefits across electrode configurations and languages, using the dataset collected by Gaddy et al. \cite{gaddy-klein-2020-digital}. This dataset contains 19 hours of 8-channel facial EMG recordings from a single English speaker during silent and vocalized speech, with a vocabulary size of 9828 words. We utilize all samples of 15 seconds or shorter in duration. In our model, we replace our Japanese mora prediction head with an English phoneme prediction head and employ the CTC loss, along with the MSE loss and the dynamic time warping loss \cite{gaddy-klein-2021-improved} between EEG/EMG latents and audio latents and constructed phoneme sequence decoders from EMG.
We trained models with the same settings as in Section~\ref{subsec:calibration_jp}. Performance was evaluated using phoneme error rate (PER).

\section{Results and discussion}
\subsection{Word classification accuracy of models trained on EEG and EMG datasets with heterogeneous configurations}
\begin{figure}[t]
  \centering
  \includegraphics[width=0.47\textwidth]{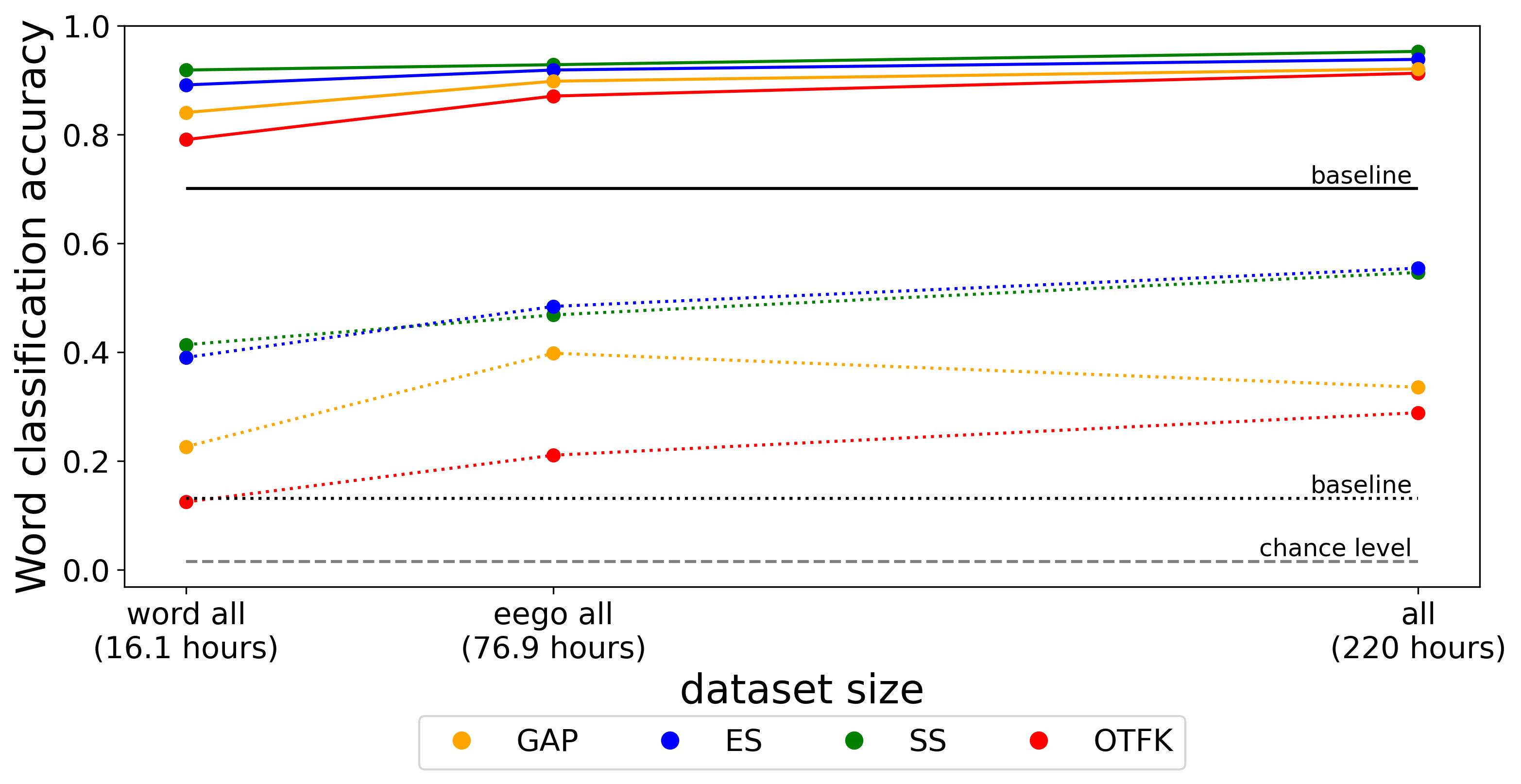}
  \caption{Classification accuracy based on dataset size, using different tokenizers. Solid lines represent mean accuracy of healthy participants (N=8), while dotted lines the patient.}
  \label{fig:all_approach_trend}
\end{figure}

Figure \ref{fig:all_approach_trend} illustrates the relationship between dataset size and performance (WCA) for models incorporating different tokenizers. All models, except for the one with GAP, showed monotonic improvements with increasing data volume, even under heterogeneous electrode configurations. This suggests these models can effectively utilize electrode-placement-agnostic information from each dataset. The model incorporating SS achieved the highest mean accuracy when trained on the ``all'' dataset, reaching $95.3\pm{2.6}$\% for healthy controls (95\% bootstrapped confidence interval) and $54.5$\% for the patient. These accuracies largely surpassed that obtained when training with each individual's data only (baseline): $70.1\pm{17.1}$\% for healthy controls and $13.2$\% for the patient. This indicates the existence of substantial shared processing across subjects and electrode configurations within the model, suggesting that minimal individual data may suffice once these are learned. However, accuracy for patient data remains notably low (though significantly higher than chance, at $1.6$\%). This can be attributed to two factors: first, the patients' mouth movements differ from those of healthy participants, extending beyond the distribution covered by healthy subject data; second, some mouth movements for different words may be fundamentally indistinguishable through EMG signals. 

\subsection{Calibration performance}

\begin{figure}[t]
  \centering
  \begin{subfigure}{0.45\textwidth}
    \centering
    \includegraphics[width=1\linewidth]{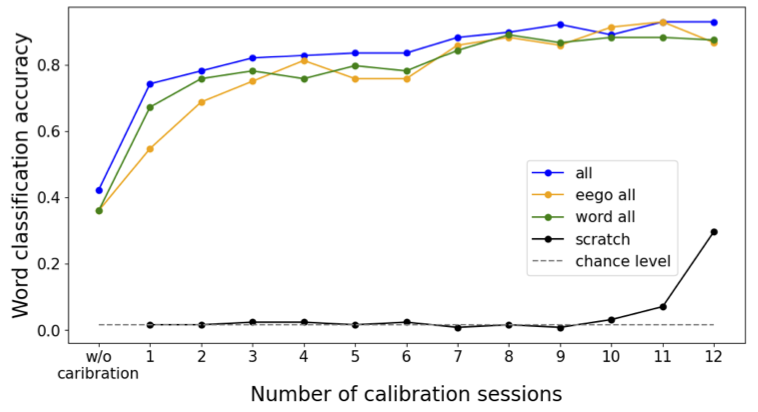}
    \caption{Healthy participant}
    \label{fig:healthy}
  \end{subfigure}
  \begin{subfigure}{0.45\textwidth}
    \centering
    \includegraphics[width=1\linewidth]{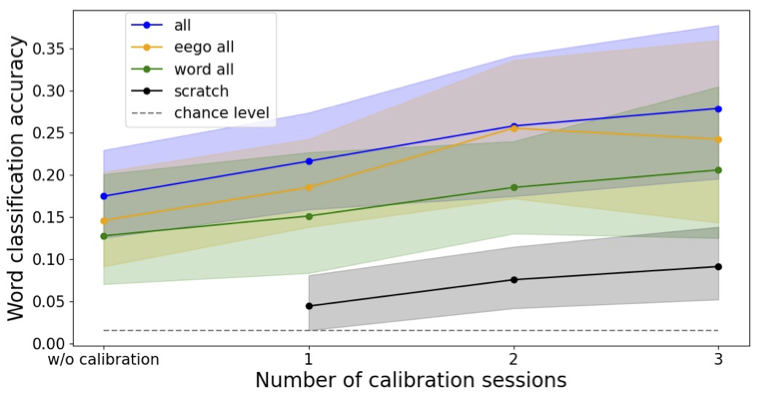}
    \caption{Patient}
    \label{fig:patient}
  \end{subfigure}
  \caption{SS model performance on unseen data with calibration sessions. Classification accuracy when calibrated for (a) a novel healthy participant and (b) new-day patient data. Solid lines and the shaded areas indicate the mean performance and 95\% bootstrapped confidence intervals.}
  \label{fig:result_calibration1}
\end{figure}
\vspace{-5pt}

Figure~\ref{fig:result_calibration1} shows calibration results for the SS model initialized with weights pretrained on the ``word all'', ``eego all'', and ``all'' datasets. Results are shown for (a) novel participant calibration and (b) new-day patient calibration. Models pretrained on the ``all'' dataset demonstrated reasonable zero-shot calibration performance (WCA), increasing to over $80$\% with just a few calibration sessions. All pretrained models consistently outperformed those trained from random initialization (scratch). These findings suggest that training with large datasets enables the models to learn processing that are robust to inter-subject and day-by-day variations inherent in biological data. However, calibration performance remained lower for patient new-day data. This can be attributed to insufficient data of patient-specific mouth movements in the pretraining data, limiting the model's ability to learn variation-invariant processing.

\begin{table}[t]
  \caption{Calibration performance (PER: lower is better) for different language and electrode configuration dataset.}
  \label{tab:gaddy_per}
  \centering
  \small
  \scalebox{0.9}{
  \begin{tabular}{ccccc}
    \toprule
    \textbf{baseline (scratch)} & \textbf{GAP} & \textbf{ES} & \textbf{SS} & \textbf{OTFK}\\ \midrule
    61.5 & 100 & \textbf{49.5} & 50.0 & 55.8\\
    \bottomrule
  \end{tabular}
  }
\end{table}

Furthermore, apart from when using the GAP tokenizer, models pretrained on the ``all'' dataset and fine-tuned on the open dataset of vocalized/silent speech in English demonstrated lower PER, as compared to a model trained from scratch (Table \ref{tab:gaddy_per}), suggesting that these models have learned some language-independent EEG/EMG processing, despite differences in Japanese as a moraic language and English as a syllabic language.

\section{Conclusion}
In this study, we construct silent speech decoders for healthy participants and a patient with a neurodegenerative disease by utilizing EEG and EMG data recorded from multiple participants with varying electrode placements. We demonstrate that training on this large-scale dataset enhances silent speech decoding accuracy despite heterogeneous electrode configurations. 
Notably, our model pretrained on large-scale data from healthy participants achieved a substantial improvement in decoding accuracy for the patient, despite the limited amount of patient-specific data available, and also demonstrated high calibration performance for a novel healthy participant, as well as new-day data for the patient. We also demonstrated transfer learning to a different dataset with a different language. Our findings not only expand the practical utility of available EEG/EMG data, but also suggest the feasibility of developing brain-computer interface systems where pretraining with large-scale data enables end users to achieve effective performance with minimal individual training data.

As a limitation, our results on transfer learning are more limited in the cross-language setting, which likely would require a multilingual pretraining dataset. In addition, due to the difficulty of acquiring patient data, we were unable to confirm how well our results generalize to other patients. In future work we plan to gather more patient data and incorporate language models to realize practical silent speech sentence decoding.

\bibliographystyle{IEEEtran}
\bibliography{mybib}

\end{document}